\documentclass[conference,letterpaper,final,10pt]{IEEEtran}
\usepackage{setspace,amssymb,graphicx,enumerate,amsfonts,amsmath,color,cite,algorithm,algpseudocode,lipsum,epsfig,subfigure,booktabs}
\usepackage[english]{babel}

\newcommand{\beq}{\begin{equation*}}
\newcommand{\eeq}{\end{equation*}}
\newcommand{\bea}{\begin{eqnarray}}
\newcommand{\eea}{\end{eqnarray}}
\newcommand{\beal}{\begin{align*}}
\newcommand{\eeal}{\end{align*}}
\newcommand{\bei}{\begin{itemize}}
\newcommand{\eei}{\end{itemize}}
\newcommand{\denk}{\begin{equation*} \begin{aligned}}
\newcommand{\denke}{\end{aligned}\end{equation*}}
\def\bary{\begin{array}}
\def\eary{\end{array}}

\begin{document}
\title{Interference-Based Cell Selection \\in Heterogenous Networks}
\author{\IEEEauthorblockN{Kemal Davaslioglu and Ender Ayanoglu}
\IEEEauthorblockA{Center for Pervasive Communications and Computing\\
Department of Electrical Engineering and Computer Science,
University of California, Irvine}}\maketitle
\author{Kemal Davaslioglu and Ender Ayanoglu}\maketitle
\begin{abstract}
Heterogeneous cellular networks provide significant improvements in terms of increased data rates and cell coverage, and offer reduced user rate starvation. However, there are important problems to be solved. In this paper, we identify that the cell selection criterion is an important factor determining the user rates especially in the uplink transmissions and apply cell breathing to determine the user and base station assignments. We observe that the proposed interference-based cell selection algorithm provides better load balancing among the base stations in the system to improve the uplink user rates. We present the implementation steps in a typical LTE network and demonstrate the performance improvements through simulations.
\end{abstract}
\IEEEpeerreviewmaketitle

\section{Introduction}
Next generation cellular systems such as Long-Term Evolution (LTE) target significantly increased throughput and capacity requirements to answer the rapidly increasing user data demand. For instance, LTE systems require a peak data rate of $100$ Mbps for downlink and $50$ Mbps for uplink in a $20$ MHz bandwidth with 64 quadrature amplitude modulation \cite{25912}. Antenna-based improvements such as employing multiple antennas at both base stations and user equipments (UEs), transmit beamforming, and spectrum-based improvements such as carrier aggregation (CA) are a few of the enabling technologies in LTE systems to achieve these challenging rates \cite{25912}. However, these will not suffice. It is obvious that, in the near future, the deployment of small cells will create a network topology shift in order to achieve significant gains that the operators need to consider along with the enabling technologies.

The deployment of different low-cost low-power base station nodes (LPN) such as picocells, femtocells and relays will provide the opportunities to increase the capacity within the macrocell area and avoid coverage holes by adapting to the varying nature of user traffic demand. According to a recent study by Ericsson, each macrocell will be overlaid with an average of three LPNs by 2017 to meet the demand for the coverage, mobility and thereby, the improved user experience \cite{ericsson}. However, to fully exploit the possible gains through heterogeneous network (HetNet) deployments, we need to consider the differences in base station types and change the conventional single-layer homogenous networks approach to include these differences. Network planning in HetNets such as LTE systems differs from conventional network planning in several aspects.

In conventional single-layer networks, base station selection was based on the highest reference signal received power (RSRP) measured at UE. While this gives the optimum selection methodology for these networks, it does not always apply to the HetNets where base stations have different transmit powers. Macrocell and picocell base stations, namely MeNBs and pico-eNBs, differ by almost $16$ dB in their downlink transmit power levels \cite{36814}. If the cell selection is based on RSRP only, UEs are more likely to connect to the MeNBs even when the path loss conditions between the pico-eNB and the UE are better. If the optimal cell selection were assigned to this case, UE could reduce its transmission power since it has higher uplink (UL) signal to interference plus noise ratio (SINR) at the closer pico-eNB, which will consequently lead to a longer battery life for the UE and reduce the interference in the system. Also, this would lead to a more balanced loading within the macrocell footprint where the resources in MeNBs and LPNs are better utilized.

In this paper, we seek to find the user-base station assignments that maximize the uplink throughput in a dense base station deployment. For this purpose, we apply the cell breathing results derived in \cite{Hanly_thesis,Hanly,Yates_BS_Selection} to the current LTE structure. We introduce an adaptive cell selection framework that enables the network to balance the traffic load between the macrocell and the picocell base stations. Our formulation enables us to exploit base station diversity to improve system performance. We compare the performance results of the proposed cell selection method to the other conventional cell selection criteria such as reference signal received power- (RSRP), cell range extension- (CRE), and path loss-based (PL) cell selection through simulations. We show that the conventional cell selection critera are not sufficient to adapt to the interference conditions when base stations are densely deployed. We observe that the interference-based cell selection criterion is better suited to reduce user rate starvation and improve median user rates in HetNet deployments with high network traffic by providing better user-base station assignments. It can provide a moderate rate increase compared to the CRE- and PL-based cell selection scheme but provides more than twice the SINR for the cell-edge users and $50\%$ improvement for the median users when compared to the RSRP-based cell selection scheme. Related works in literature include \cite{energyefficiency} that also incorporates cell breathing to a macrocell-femtocell network with a frequency reuse of three. Another related work is \cite{qualcomm} where similar results comparing RSRP- and CRE-based cell selection are shown. This work differs from \cite{qualcomm} in that we employ interference-based cell selection, and the distinction between the work in \cite{energyefficiency} and here will be presented in the sequel.

\section{System Basics}
In this section, we introduce the system basics and the nomenclature used in this paper. We will briefly introduce the transmission schemes and the uplink power control proposed in LTE standards. Uplink transmissions in LTE are designed to be highly power-efficient to improve the coverage and to reduce the power consumption at UE \cite{gatech}. For this purpose, single carrier frequency-division multiple access (SC-FDMA) is adopted \cite{25912}. SC-FDMA enables a smaller peak-to-average power ratio compared to the regular orthogonal frequency division multiplexing access (OFDMA) \cite{holma}. In this paper, we use localized FDMA where the user outputs are mapped to consecutive subcarriers as defined in \cite{25912}.

\begin{figure}[tb!]\centering
  \includegraphics[width=3in]{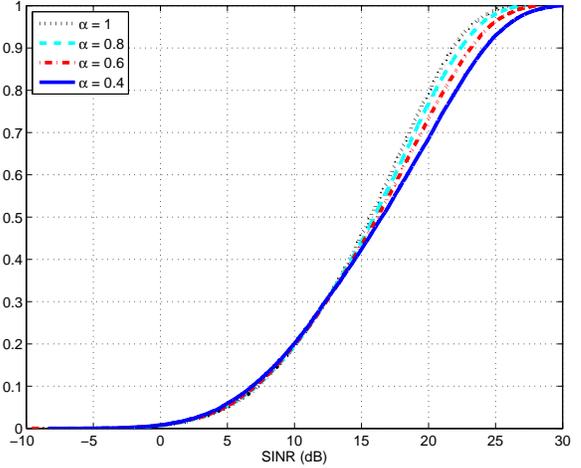}
 \caption{The effects of various path loss compensation factors, varying from fractional path loss compensation values $\alpha = \{0.4, 0.6, 0.8 \}$ to the full path loss compensation of $\alpha = 1$, on the c.d.f. of user SINR in a heterogenous network deployment with $2$ picocells per sector. $P_0$ is taken as $-90$ dBm and the proposed interference-based cell selection is used for user-base station assignments.\label{alphaFairnessCdf}}
\end{figure}

We consider a system with $K$ users and $B$ base stations, and let $c_k$ denote the base station that user $k$ is associated with. We form an $K\times 1$ vector $\textbf{c}$ to represent all the user-base station assignments in the system. The LTE specification in \cite{36213} defines the closed-loop uplink power control as
\begin{align}\label{pcClosedLoop} P_k = \min\{P_{\max},P_0 + 10 \log_{10}(N_{RB_k}) &+ \alpha PL_{k,c_k} \\&+ \Delta_{TF_k} + f_k\} \nonumber \end{align}
where $P_k$ denotes the uplink transmit power of user $k$ and represented in dBm units above, $P_{\max}$ denotes the maximum UE transmit power, and $P_0$ is open loop transmit power. $N_{RB_k}$ represents the number of resource blocks (RBs) that are assigned to user $k$, and $PL_{k,c_k}$ is the path loss between user $k$ and its serving base station $c_k$. Note that the path loss also includes the shadow fading of the link. The path loss compensation factor $\alpha$ takes its value from the following set, $\alpha \in \{0,0.4,0.5,0.6,0.7,0.8,0.9,1\}$. $\Delta_{TF_i}$ is a parameter based on the modulation and coding scheme and $f_i$ is a closed loop power control adjustment parameter. It is important to emphasize that path loss compensation factor $\alpha$ determines the fairness within the cell such that for a full path loss compensation, $\alpha = 1$, the network enables the cell edge users, that have high path loss values, to transmit at high power levels. The fairness in the system decreases as the path loss compensation approaches zero, i.e., $\alpha \rightarrow 0$ since the high path loss for the cell-edge users are not compensated for. Consequently, this improves the rates for the cell center and median users due to less interference compared to the full path loss compensation case, i.e., $\alpha = 1$. Fig. \ref{alphaFairnessCdf} shows the effects of path loss compensation factor $\alpha$, on user SINR cumulative distribution function (c.d.f.) in a heterogenous cellular network. We note that Fig. \ref{alphaFairnessCdf} assumes the same simulation setup and parameters presented in Section \ref{simulationSection}.

In this paper, we will investigate the performance of the users under open loop power control. The user power, in units of $\mathrm{dBm}$, is given by
\begin{align} P_k = \min\{P_{\max},P_0 + 10 \log_{10}(N_{RB_k}) + \alpha PL_{k,c_k}\} \end{align}
where $\Delta_{TF_k}$ and $f_k$ in (\ref{pcClosedLoop}) are ignored in the system level simulations as in \cite[pp. 195]{holma}. The UE transmission power is equally distributed on the allocated bandwidth. The corresponding UE transmit power spectral density in $\mathrm{dBm/Hz}$ is given by
\begin{align} P_{k,n} = P_0 + \alpha PL_{k,c_k} \end{align}
where $P_{k,n}$ denotes the uplink transmit power of user $k$ on subcarrier $n$.

\section{Cell Selection Techniques}
\subsection{Conventional Cell Selection Schemes}
The RSRP-based cell selection is carried out by choosing the base station that maximizes the received power of the reference signals such that
\begin{align} c_k = \arg \max_{j \in \mathcal{B}}\hspace{.05in} g_{jk}\widetilde{p}_j = \arg \max_{j \in \mathcal{B}} \hspace{.05in} \mathrm{RSRP}_j \label{RSRPCellSelection}
\end{align}
where $\mathcal{B}$ is the base station search space, $g_{jk}$ is the channel gain between base station $j$ and user $k$, and  $\widetilde{p}_j$ denotes the average power of the cell-specific downlink reference signals for the $j$th base station within the considered measurement bandwidth as they are defined in the standards \cite{36214}.

The key point in our analysis is that we assume the average power allocated to the reference signals by MeNBs, $\widetilde{p}^{MeNB}_m$, are larger compared to those of pico-eNBs, $\widetilde{p}^{peNB}_p$, i.e., $\widetilde{p}^{MeNB}_m > \widetilde{p}^{peNB}_p, \forall m \in \mathcal{M}, \forall p \in \mathcal{P}$ where $\mathcal{M}$ and $\mathcal{P}$ denote the sets of macrocell and picocell base stations such that $\mathcal{M} \cup \mathcal{P} = \mathcal{B}$. The reason for this assumption is that the coverage area of an MeNB, $\mathcal{A}_{MeNB}$, is significantly larger than the coverage area of a pico-eNB, $\mathcal{A}_{peNB}$, and in order to provide full coverage within the cell, more power needs to be allocated for the MeNB reference symbols, i.e.,
$\mathcal{A}_{MeNB} > \mathcal{A}_{peNB}\Rightarrow \widetilde{p}^{MeNB}_m > \widetilde{p}^{peNB}_p,\forall m \in \mathcal{M}, \forall p \in \mathcal{P}$.

\begin{figure}[tb!]\centering
  \includegraphics[width=3in]{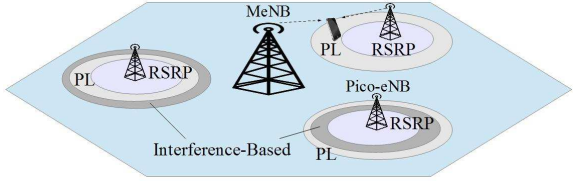}
 \caption{Difference in coverage regions of picocells under RSRP-based, PL-based and interference-based cell selection strategies in a HetNet layout.\label{CoverageAreas}}
\end{figure}

For PL-based cell selection, user $k$ selects the base station with the maximum channel gain based on its reference signal measurements such that \begin{align} c_k = \arg \max_{j \in \mathcal{B}}\hspace{.05in} g_{jk}. \end{align}
An important observation is that, based on these definitions, it is straightforward to see that the coverage area of a picocell with PL-based cell selection is always greater than its coverage area with RSRP-based cell selection in a HetNet layout such that
\begin{align} \mathcal{A}_{peNB_p}^{PL} \geq \mathcal{A}_{peNB_p}^{RSRP}, \hspace{.1in} \forall p \in \mathcal{P}.\end{align}
We depict this result in Fig. \ref{CoverageAreas}. Also, note that in single layer homogenous networks, RSRP- and PL-based cell selections are equivalent since base stations allocate the same power levels for reference signals. Although a variation of PL-based cell selection was employed in Global System for Mobile (GSM) \cite{GSM3gpp}, this criterion is not included in the LTE standards. However, we include it as a reference in our paper since it potentially provides increased uplink rates by better user-base station assignments compared to RSRP-based cell selection.

The third cell selection method we investigate is to use the cell range expansion to extend the picocell coverage with a constant offset to offload macrocell users to picocells. Similar to the RSRP-based cell selection, this criterion also uses the RSRP measurements at the UE for cell selection. CRE-based cell selection is expressed as
\begin{align} c_k = \arg \max_{j \in \mathcal{B}} \hspace{.05in} \mathrm{RSRP}_j^{dB} + \mathrm{Bias}_j^{dB}\end{align}
where the above terms are expressed in dB, $\mathrm{Bias}_j = 0$ for all macrocell base stations, $m \in \mathcal{M}$, and typical values for picocell base station offset values are $3$, $6$, $9$, and $12$ dB. When we consider a user $k$, a macrocell base station $m \in \mathcal{M}$ and a picocell base station $p \in \mathcal{P}$, user $k$ chooses the picocell base station $p$ with a $\mathrm{Bias}_p$ under CRE-based cell selection if the following holds true
\begin{align} \mathrm{RSRP}_m^{dB} < \mathrm{RSRP}_p^{dB} + \mathrm{Bias}_p^{dB} \label{CreRsrpFormula}\end{align}
where a large offset $\mathrm{Bias}_p$ increases the picocell coverage. Although this may offer increased rates in the uplink, the opposite is true for the downlink user rates. Especially, the users in the range expansion area are exposed to severe cross-tier interference from macrocell base stations in the downlink. Therefore, the offset values and consequent picocell coverage regions need to be carefully adjusted such that the imbalance in the downlink and the uplink user rates can be mitigated.

The main drawback of these three cell selection schemes is that they solely depend on path loss and downlink transmit powers and do not consider the instantaneous interference in the system. In the case of high interference at a base station either caused by increased number of interfering users in neighboring cells or interfering users with high data rate demand, a better approach is to hand over excess users to neighboring cells to better balance the load among the base stations. The following section introduces an adaptive cell selection criterion that mitigates this drawback.

\subsection{Interference-Based Cell Selection}
Let $n$ be a subchannel assigned to user $k \in \mathcal{U}$ and $\Theta_k$ denote the set of subcarriers within the consecutive RBs assigned to user $k$. The uplink interference plus noise experienced by the serving base station $S$ on the set of subcarriers $n \in \Theta_k$ can be represented as
\begin{align} I_{S} = \sum_{n \in \Theta_k} \left( \sum_{u \in \mathcal{U},\hspace{.01in} u\neq k}\hspace{.01in} p_{u,n} \hspace{.025in} g_{u,S}+ \sigma^2_S \right) \label{iserving}\end{align}
where $p_{u,n}$ denotes the transmit power of the interfering users $u \in \mathcal{U}$ on subcarrier $n$, and $g_{u,S}$ denotes the channel gain between the user $u$ and the base station $S$. Note that the users that are associated with the same base station are orthogonal in the frequency domain, and the interfering users are the users that are allocated to the same resource blocks in the neighboring cells. On the other hand, the interference plus noise at a candidate base station $C$ on the set of subcarriers in the resource blocks $n \in \Theta_k$ is given by
\begin{align} I_{C} = \sum_{n \in \Theta_k} \left(\sum_{u \in \mathcal{U}}\hspace{.05in} p_{u,n} \hspace{.025in} g_{u,C}+\sigma^2_C \right) \label{icandidate}\end{align}
where the uplink transmission of user $k$ is also considered as an interfering link to the candidate cell $C$ since user $k$ is not served by the candidate base station C. In (\ref{RSRPCellSelection}), we have shown that the downlink RSRP measurements can be used in cell selection procedure. When the uplink and downlink channel gains are symmetric, typically a valid assumption for time domain duplexing systems, downlink reference signal (RS) broadcasts can also be used to estimate the uplink channel gains such that $g_{u,c} \approx g_{c,u}$. The proposed cell selection method can be used in a frequency domain duplexing system based on the sounding reference signals (SRS) transmitted by UEs to estimate the uplink channel quality at the base station. In either case, we can rewrite the cell selection rule between the serving and the candidate base stations. User $k$ selects the candidate base station $C$ if the following is true
\begin{align} \frac{\widetilde{p}_{S} \hspace{.01in} I_S}{\mathrm{RSRP}_S} > \frac{\widetilde{p}_{C} \hspace{.01in} \left(I_C - \sum_{n \in \Theta_k} p_{u,n} g_{u,C} \right)}{\mathrm{RSRP}_C} \label{Icomparison}\end{align}
where $\widetilde{p}_S$ and $\widetilde{p}_C$ denote the cell-specific downlink reference signal broadcast from the serving and candidate cells, respectively, and we used the fact that $g_{u,C} = \mathrm{RSRP}_C/\widetilde{p}_C$ to derive (\ref{Icomparison}). Note that the interference caused by uplink transmissions of user $k$ is subtracted from the total interference on the candidate cell in (\ref{Icomparison}). When we rearrange terms, we obtain
\begin{align} \mathrm{RSRP}_C > \mathrm{RSRP}_S \hspace{.015in} \frac{\widetilde{p}_{C}}{\widetilde{p}_{S}} \hspace{.01in} \frac{\left(I_C - \sum_{n \in \Theta_k} p_{u,n} \hspace{.01in}g_{u,C} \right)}{I_S}.\end{align}
Notice that the above form is very similar to the CRE-based cell selection in (\ref{CreRsrpFormula}). Rather than applying a predetermined constant bias value, we can now obtain an adaptive offset for each base station for different resource blocks when interference-based cell selection is used. Thus, the interference-based bias value is given by
\begin{align} \mathrm{Bias}_j = \frac{\widetilde{p}_{C}}{\widetilde{p}_{S}} \hspace{.01in} \frac{\left(I_C - \sum_{n \in \Theta_k} p_{u,n}\hspace{.01in} g_{u,C} \right)}{I_S}.\end{align}
Note that this approach assumes that user $k$ is scheduled the same resource blocks in both its serving cell and its candidate cell.

Hence, to summarize our results in this section, the general rule of interference-based cell selection for user $k$ can be expressed as
\begin{align} c_k^* = \arg \min_{c \in \mathcal{B}} \sum_{n \in \Theta_k} \frac{\sum_{u \in U, \hspace{.01in} u \neq k}\hspace{.05in} p_{u,n} \hspace{.025in} g_{u,c} + \sigma_c^2}{g_{k,c}}. \end{align}

We observe that the coverage area of a picocell with interference-based cell selection depends on the interference in each cell. It can be seen that, under heavy traffic, the coverage area of our algorithm is smaller than both RSRP-based and PL-based cell selection criteria. A smaller cell size is desirable under heavy traffic, such that extra traffic can be offloaded to the neighboring cells. On the other hand, the proposed cell selection method results in a larger cell size than both RSRP-based and PL-based cell selection schemes under light load, and it is indeed desirable to have larger cells to extend coverage under light load.

Our proposed method, although based on \cite{energyefficiency}, differs from \cite{energyefficiency} in several aspects. First, the approach in \cite{energyefficiency} employs a macrocell-femtocell layout in which macrocells are allowed to operate at a single band and femtocells can select one of the three subbands that only one overlaps with the band that macrocell operates at. Hence, the model in \cite{energyefficiency} inherently applies a frequency reuse of three among base stations. In our paper, we do not impose this constraint. Instead, we apply a full bandwidth share that the macrocells and picocells can operate at any subband within the channel bandwidth. We consider the interference on each resource block and adjust the cell coverage accordingly. Thereby, our work considers a more efficient and realistic model considering LTE standards. Second remark is that we focus on the SINR and rate improvements whereas \cite{energyefficiency} considers the power consumption and handover probability.
\section{Simulation Results}\label{simulationSection}
\begin{figure}[tb!]\centering
  \includegraphics[width=2.575in]{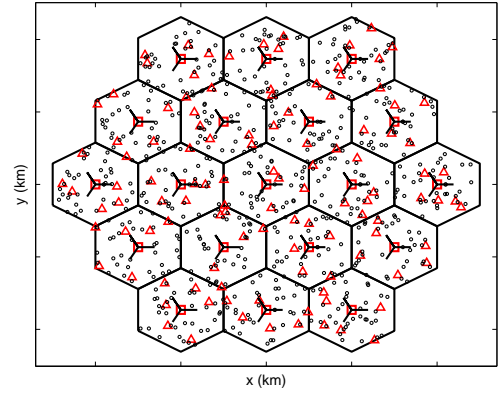}
 \caption{Figure depicts the simulation layout of a HetNet deployment. The layout includes an idealized $19$ macrocells with each employing a 3 sector antenna that is overlaid with $2$ pico-eNBs per sector. The simulation is carried out for $12$ active users per sector. Macro- and picocell base stations and users are represented by squares, triangles and circles, respectively.}\label{layout}
\end{figure}

In this section, we present the performance results of the proposed cell selection method and compare its performance to RSRP-, PL- and CRE-based cell selection strategies in a HetNet deployment scenario. As a measure of performance, we focus on the $90$th, $50$th and $5$th percentile users. These typically correspond to cell center, median and the cell-edge user rates, and the latter two are also defined in standards such as \cite{36814}.

\begin{table}[!t]
\caption{Simulation Parameters}
\label{sim_params}
\centering
\small{\begin{tabular}{cc}\toprule
\textbf{Parameter} & \textbf{Setting} \\\midrule
UE to MeNB channel model & $128.1+37.6\log_{10}(d)$\\
UE to Pico-eNB channel model & $140.7+36.7\log_{10}(d)$ \\
Inter-site distance & $500$m\\
Minimum Pico- to Macro- distance & $75$m\\
Minimum Pico- to Pico- distance & $35$m\\
Total number of Data RBs  & $48$ RBs \\
Number of RBs for each user & 4 RBs\\
Maximum UE Power & $23$ dBm \\
Effective Thermal Noise Power & $174$ dBm$/$Hz \\
Noise Figure, Total Bandwidth& $5$dB, $10$ MHz \\
MeNB Rx Antenna Gain   & $15$ dB\\
Pico-eNB Rx Antenna Gain & $5$ dB\\
Antenna Horizontal Pattern, $A(\theta)$ & $-\min(12(\theta/\theta_{\mathrm{3dB}})^2,A_m)$\\
$A_m$, $\theta_{\mathrm{3dB}}$  & $20$ dB, $70^\circ$\\
Penetration Loss & $20$ dB \\
UE to MeNB Shadowing & $\sigma = 8$ dB\\
UE to Pico-eNB Shadowing & $\sigma = 10$ dB\\\bottomrule
\end{tabular}}
\end{table}

The simulation scenario we consider in this section is shown in Fig. \ref{layout}. It assumes a HetNet deployment of idealized $19$-cell macrocells, each employed with $3$-sector antennas, and they are shown with squares. In our simulations, we investigated the performance improvements achieved by moderate and dense picocell deployments. For this purpose, we first simulated two randomly placed picocell base stations in each sector. Then, we repeated our simulations for a dense deployment of six picocells per sector. We initially placed one user per picocell within the picocell coverage of $50$ m and randomly placed the remaining users within the macrocell sector. A total of $12$ users per each sector are investigated. The purpose of this type of user distribution is to observe the under utilization of picocell base station overlay and to see the improvements offered by the proposed cell selection method. The minimum pico-eNB to pico-eNB and pico-eNB to MeNB distances are shown in Table \ref{sim_params} along with the other system simulation parameters. These parameters and channel models are taken from the proposed models in \cite{36814} for the heterogeneous system simulation baseline parameters. The downlink reference signal powers are taken as $\widetilde{p}_{MeNB} = 46$ dBm and $\widetilde{p}_{peNB} = 30$ dBm. We assumed a full buffer model for the users where each user has an infinite queue length and always has data to transmit. The wrap around technique is used to avoid edge effects. Also, we assume that minimum mean square error (MMSE) equalizers are employed at the base station receivers. Then, the wideband SINR for each user $\gamma_k$ can be obtained using the SINR of each $N_{RB_k}$ subcarriers assigned to user $k$, $\gamma_{k,n}$ as \cite{goodman}
\begin{align} \gamma_k = \left(\frac{1}{\frac{1}{N_{RB_k}}\sum_{n=1}^{N_{RB_k}}\frac{\gamma_{k,n}}{\gamma_{k,n}+1}}-1 \right)^{-1}. \end{align}

\begin{figure*}[h!]\centering
\begin{tabular}{cc}
\subfigure[]{\includegraphics[height=2in,width= 0.45\linewidth]{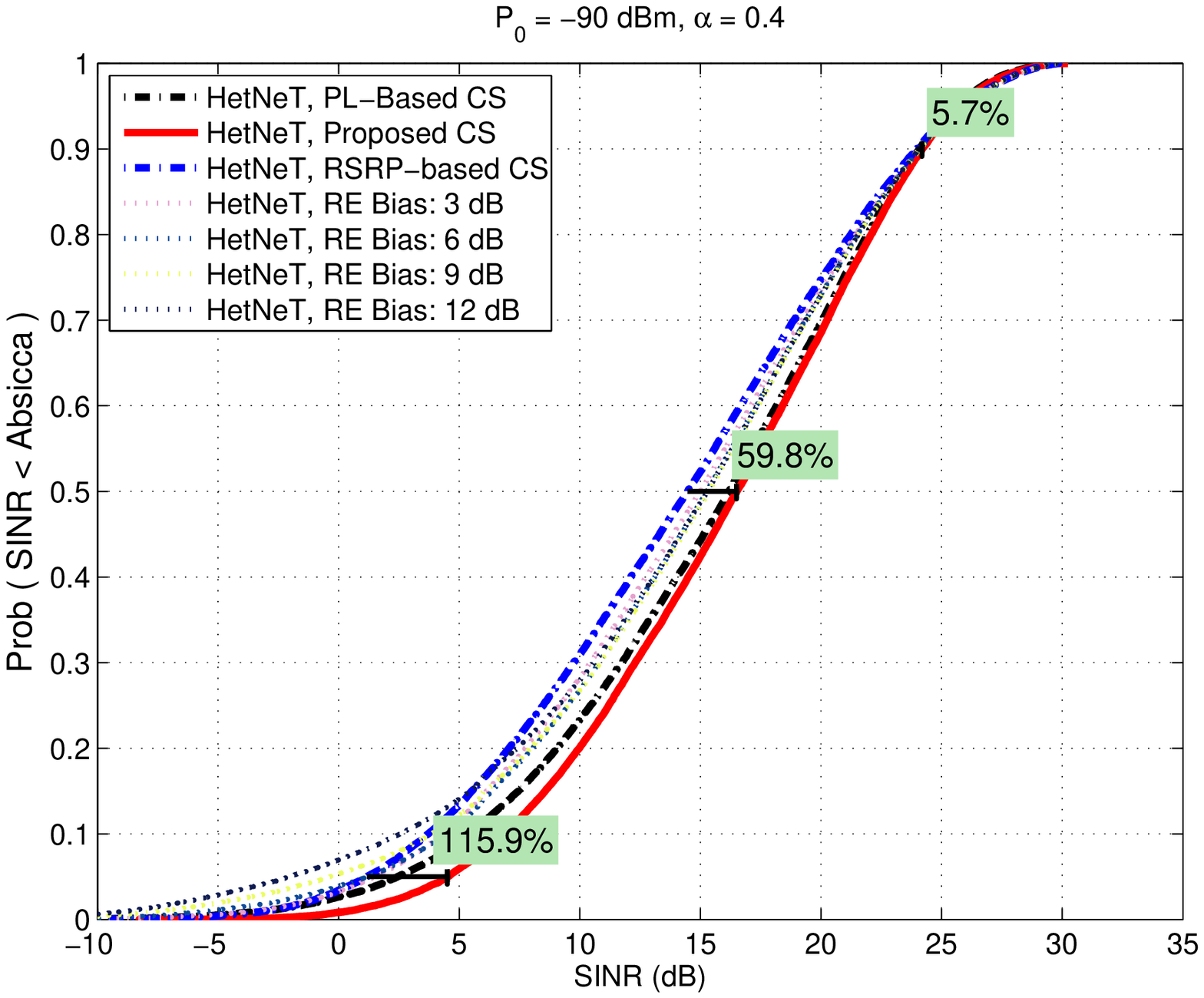}}
\subfigure[]{\includegraphics[height=2in,width= 0.45\linewidth]{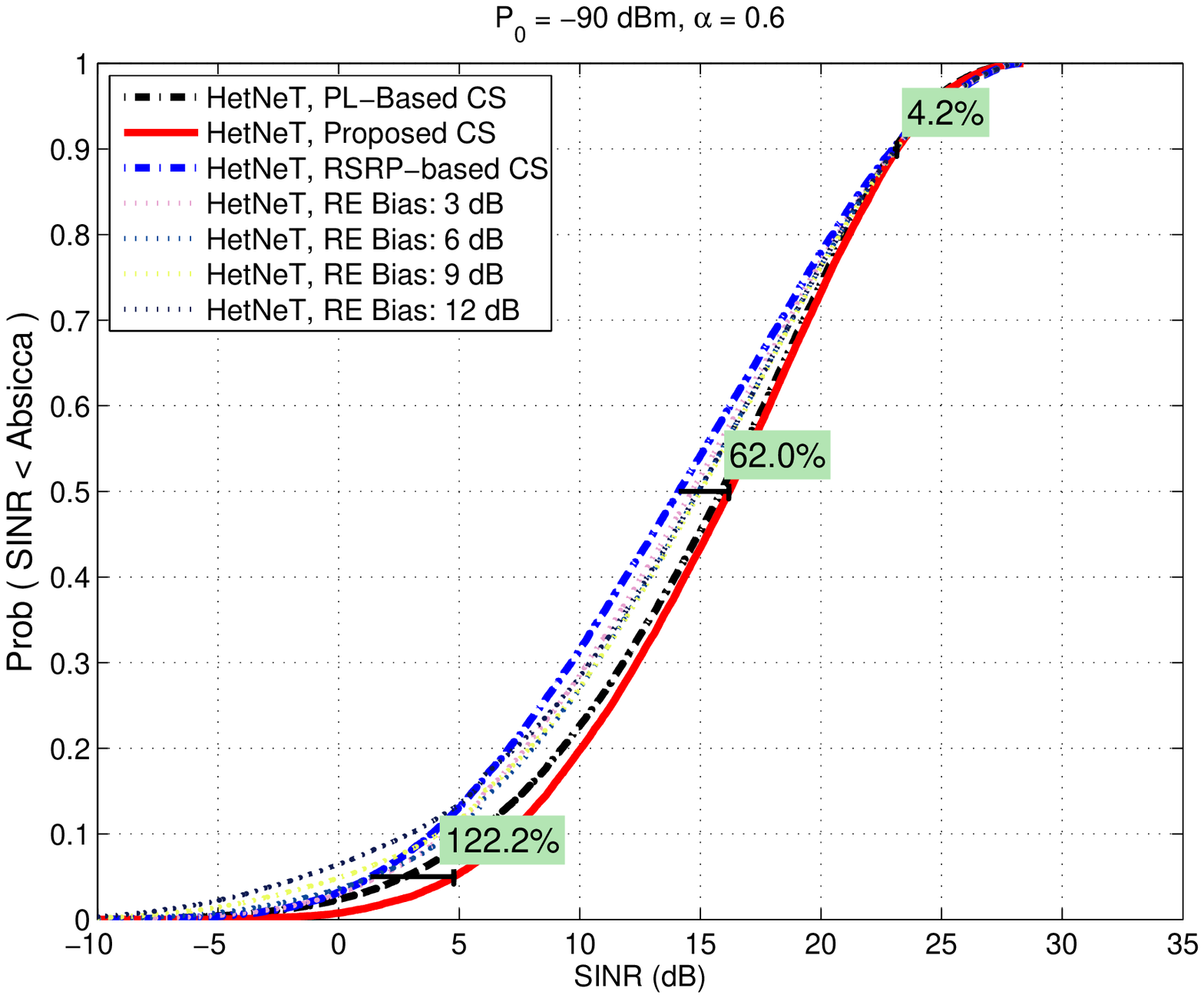}}\\
\subfigure[]{\includegraphics[height=2in,width= 0.45\linewidth]{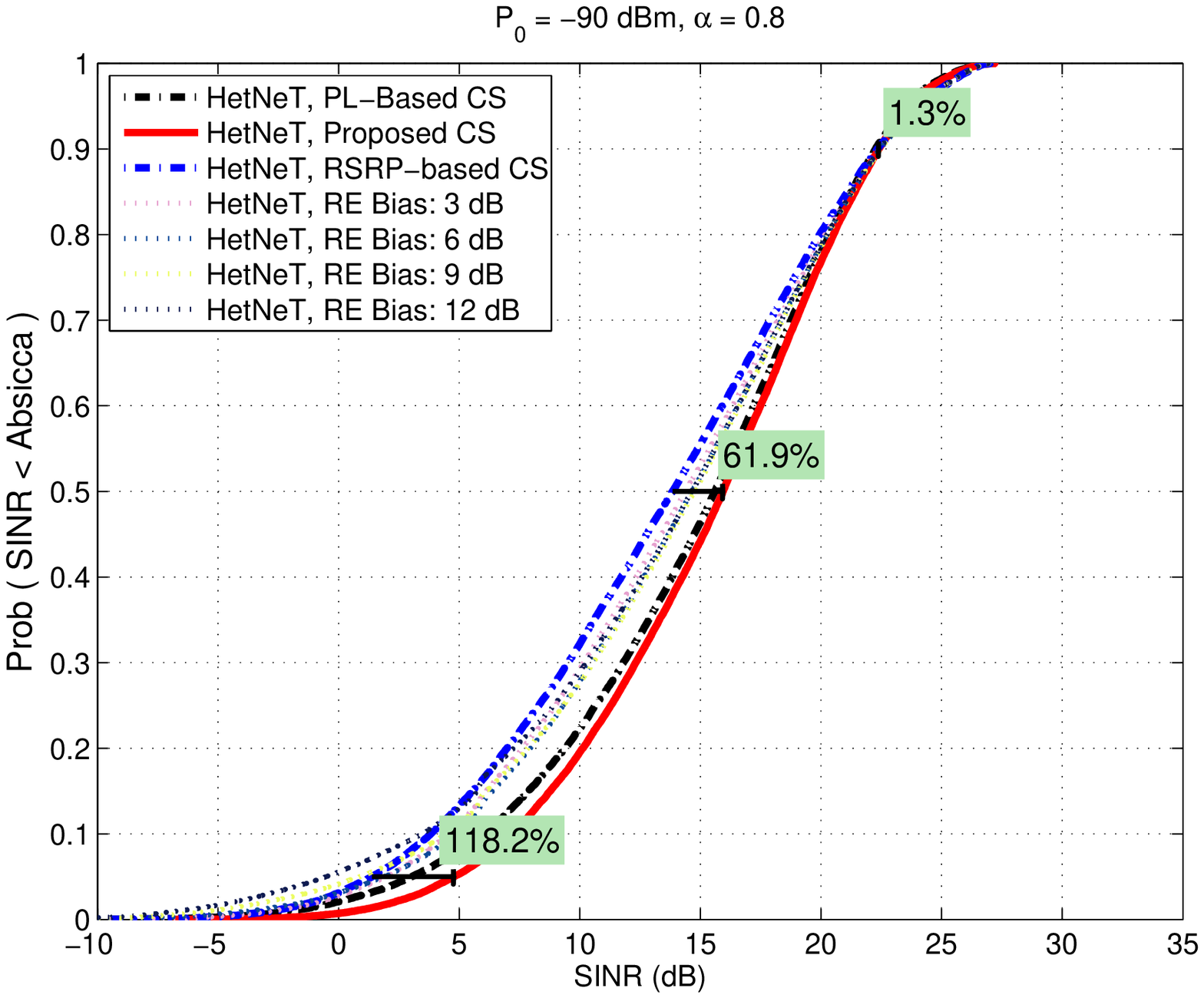}}
\subfigure[]{\includegraphics[height=2in,width= 0.45\linewidth]{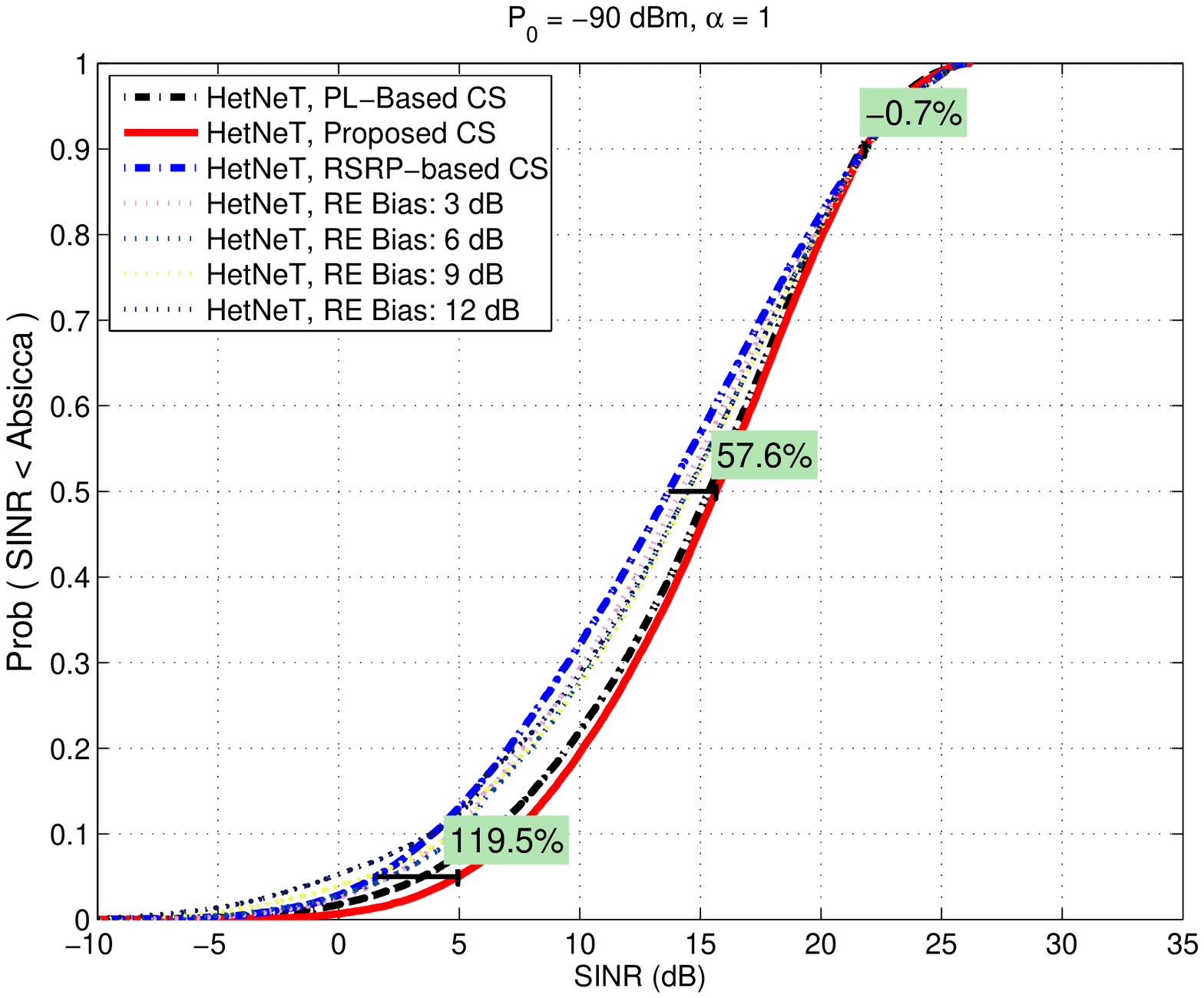}}
 \end{tabular}
 \caption{Each plot displays the c.d.f. of of user wideband SINR with different cell selection strategies for $P_0 = -90$ dBm and $\alpha$ varying between $0.4$ to $1$ (fractional PL compensation to full PL compensation) in a HetNet deployment of $2$ picocells and $12$ users per sector. The marked percentiles of $5$, $50$ and $90$th percentile represent cell-edge, median and cell center users, respectively.}\label{2picos}
 \begin{tabular}{cc}
\subfigure[]{\includegraphics[height=2in,width= 0.45\linewidth]{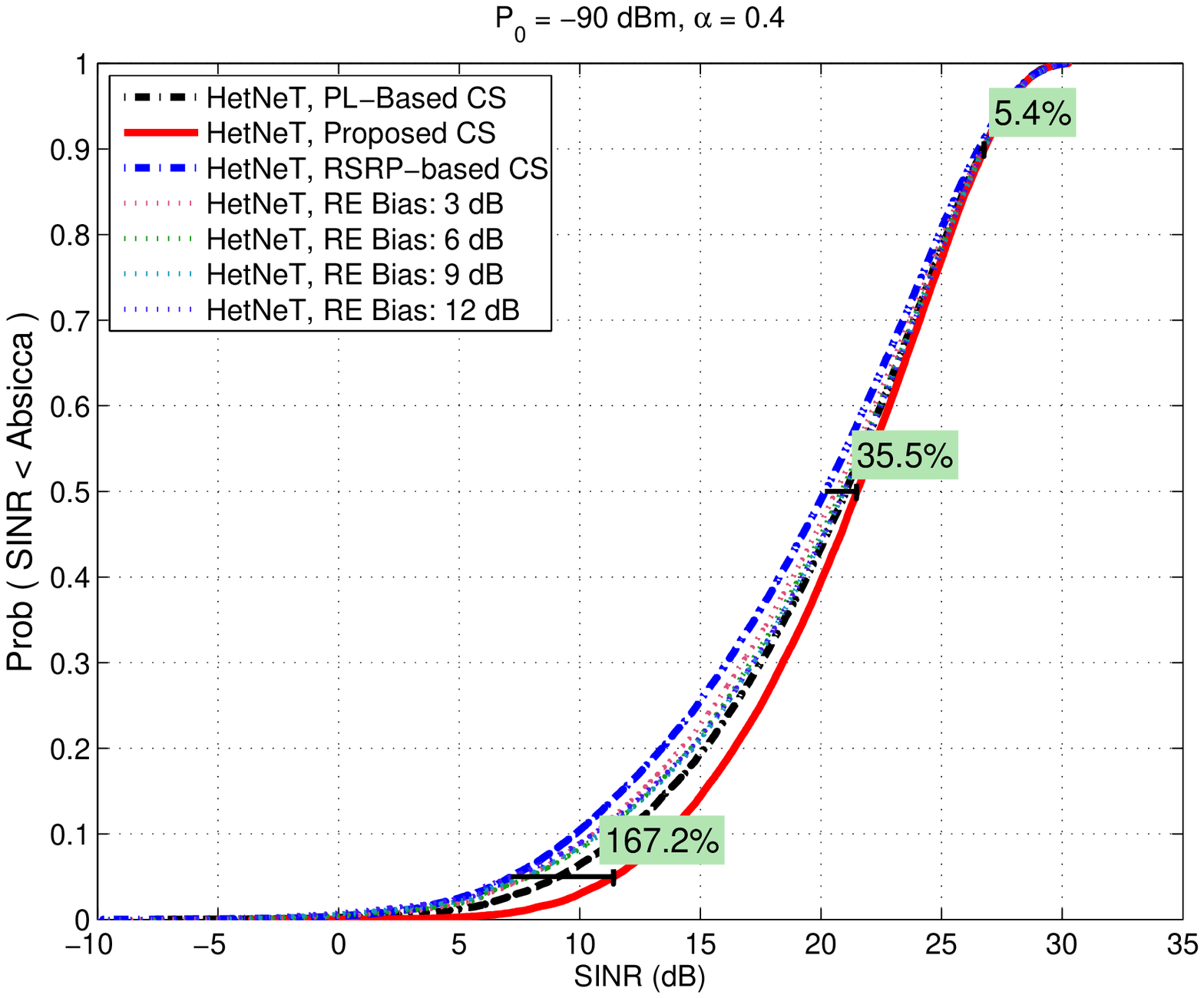}}
\subfigure[]{\includegraphics[height=2in,width= 0.45\linewidth]{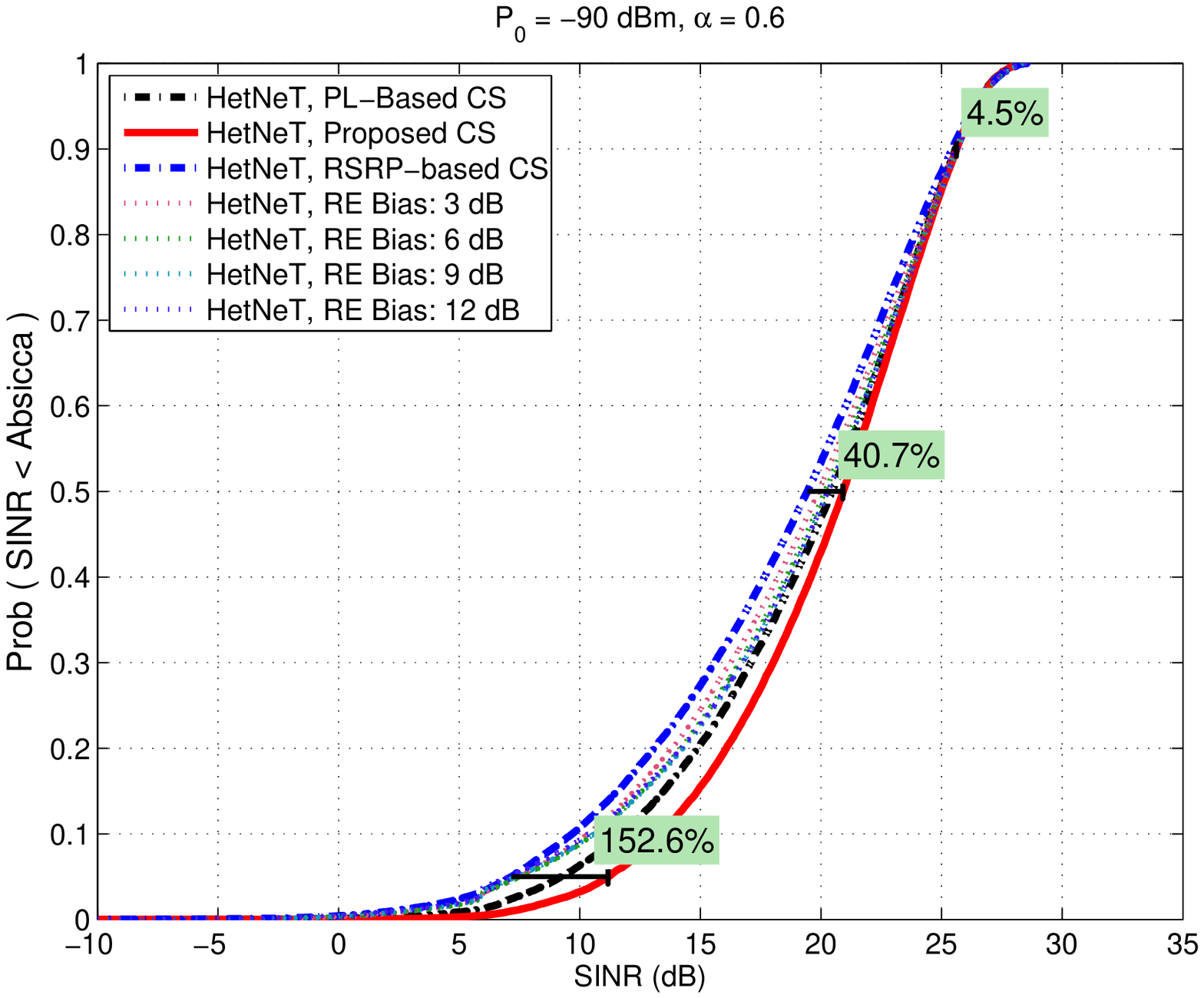}}\\
\subfigure[]{\includegraphics[height=2in,width= 0.45\linewidth]{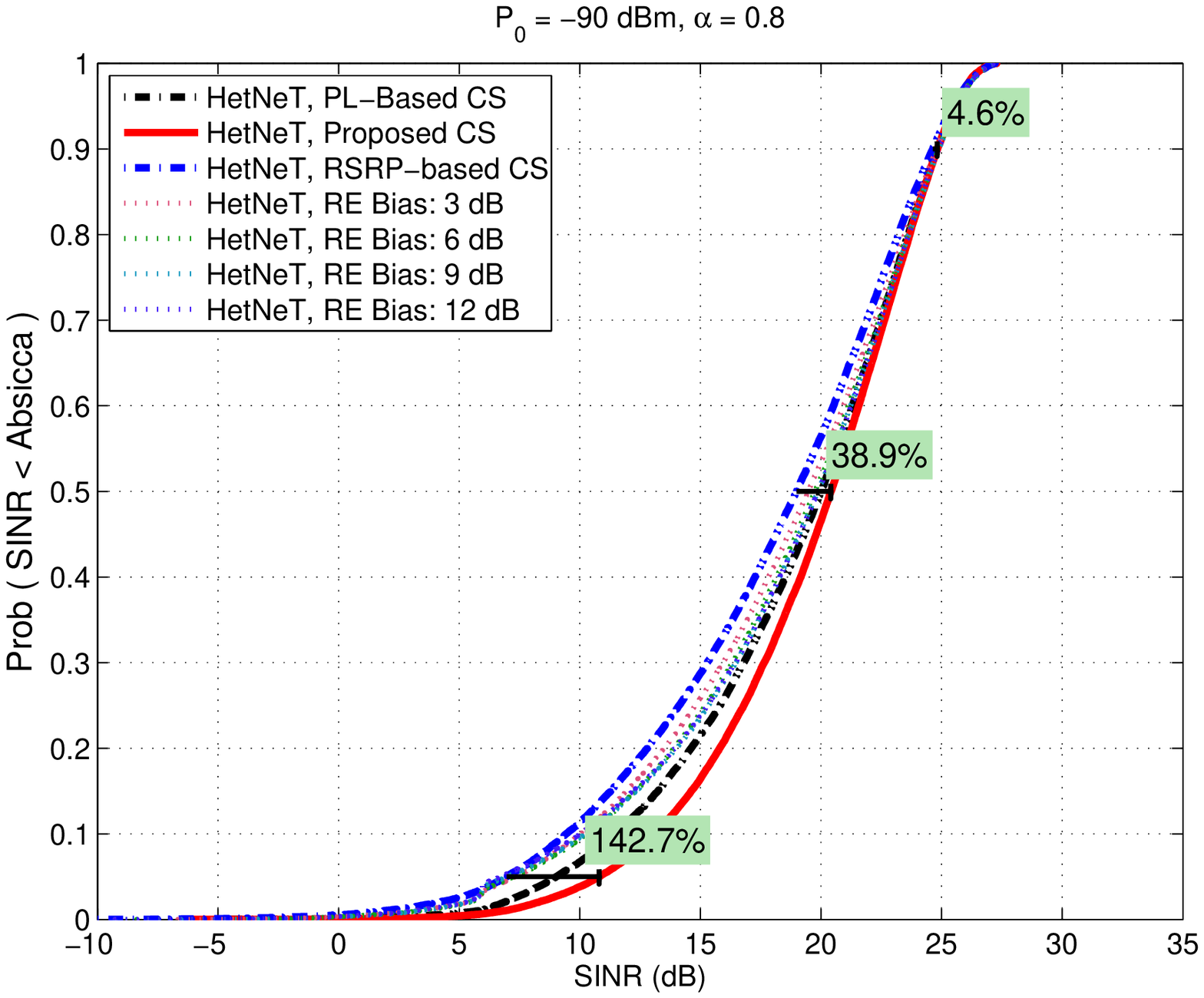}}
\subfigure[]{\includegraphics[height=2in,width= 0.45\linewidth]{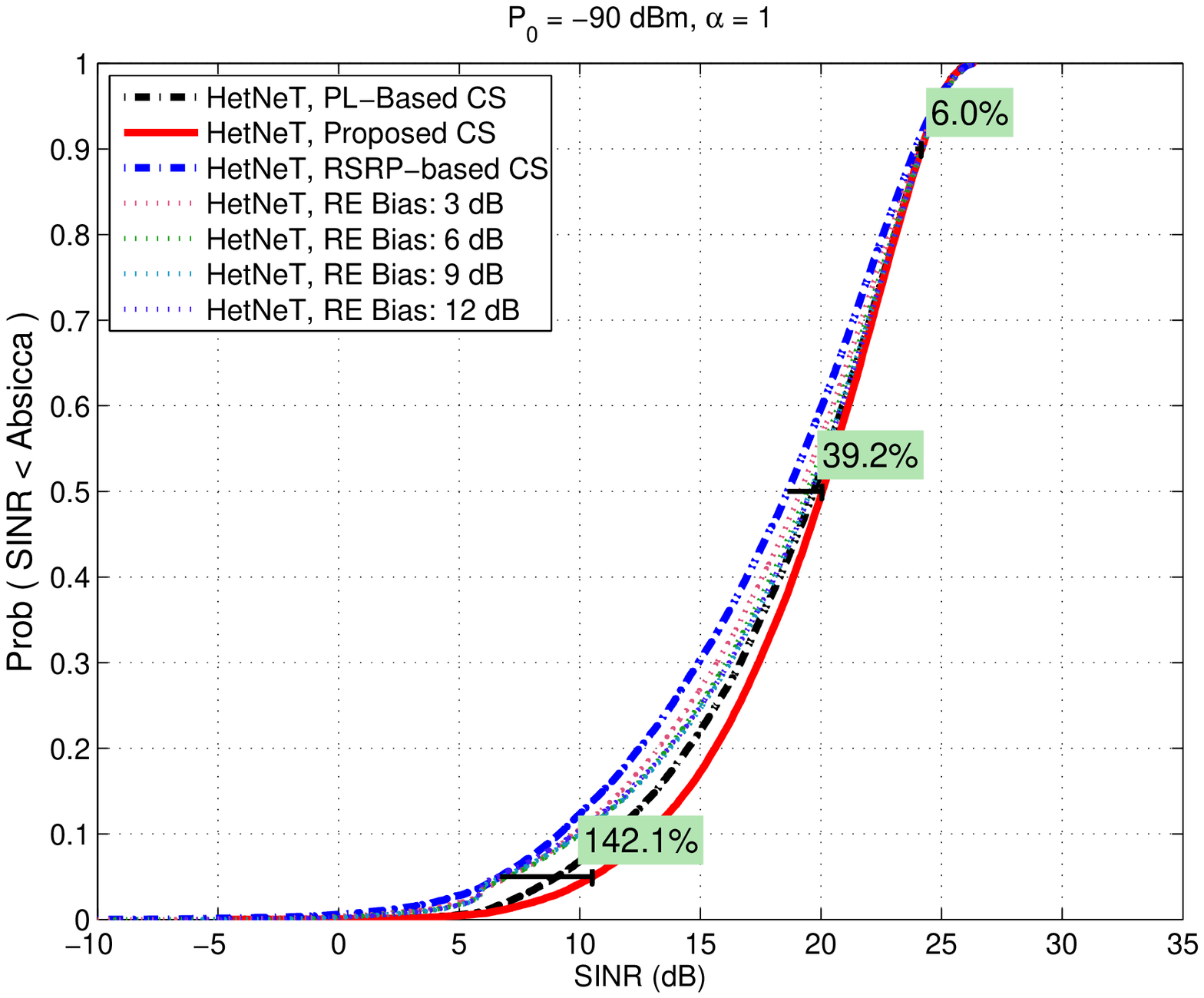}}
 \end{tabular}
 \caption{Each plot displays the c.d.f. of of user wideband SINR with different cell selection strategies for $P_0 = -90$ dBm and $\alpha$ varying between $0.4$ to $1$ (fractional PL compensation to full PL compensation) in a HetNet deployment of $6$ picocells and $12$ users per sector.}\label{6picos}
\end{figure*}

Fig. \ref{2picos} (a)-(d) displays the c.d.f. of user wideband SINR for different $(P_0,\alpha)$ pairs. We assumed a moderate picocell deployment of $2$ picocells per sector. We see that for all three types of users, cell center, median and cell-edge users, the proposed interference-based cell selection outperforms the conventional cell selection methods of RSRP-, CRE- and PL-based cell selection. The marks in the figure are to compare the RSRP- and interference-based cell selection methods. We see that cell-edge users experience the most significant SINR improvements of achieving more than double gains in SINR with the proposed scheme. These gains are achieved through adjusting to the interference conditions at each base station. For instance, if a base station has high interference from its neighboring cells, it is often advantageous to offload some of its users to the other cells. This type of allocation enables the cell-edge users or the users with bad channel conditions to significantly improve their rates. We note here that this type of cell selection scheme performs best when base stations have overlapping cells. In the case where base stations are sparsely deployed, these gains may not be achieved.

Next, we investigate a heterogenous network deployment with a dense small cell overlay. We simulate the previous system setup with $6$ picocells per sector. The resulting c.d.f. versus SINR results are depicted in Fig \ref{6picos} (a)-(d). As a first note, we can see a significant SINR improvement achieved by the dense base station deployment when we compare Figs. \ref{2picos} and \ref{6picos}. We observed more than $5$ dB gains in SINR can be achieved when we increase the number of picocells per sector from two to six. In fact, these gains are mainly obtained through finding either closer base stations or base stations with less interference. Hence, these improvements are the direct results of increased base station diversity. Obviously, in terms of network operators, this comes with additional capital and operational expenses for the dense picocell deployments. However, we can see that this translates into direct gains in terms of SINR. Another observation we can make based on Fig. \ref{6picos} (a)-(d) is that the previous results for the $2$ picocells per sector deployments still apply to the $6$ picocells per sector. With the interference-based cell selection method, the SINR offered to the cell-edge users or users high path loss values are significantly improved compared to RSRP-, CRE- and PL-based cell selection methods. Likewise, the proposed cell selection method also provides increases in user SINR around $30-40\%$ for the median users.

\section{Conclusion}
The deployment of heterogeneous base stations provides substantial gains on the cellular network performance in terms of increased data rates, improvements in cell coverage and significantly reduced user outages. In order to fully utilize the benefits from heterogenous base station deployments, a different approach in network planning needs to be pursued and for this purpose, we identified the critical role of the cell selection criterion for the user-base station assignments. Conventional cell selection schemes such as RSRP-, CRE- and PL-based strategies ignore the network traffic load and often times do not provide the optimal solution. Instead, interference-based cell selection offers more flexibility to adapt to the varying traffic load and user mobility. Based on our simulation results, we conclude that the proposed interference-based cell selection provides significant improvements in the uplink to double the cell-edge user SINR and increase the median user SINR by $50\%$ when compared to the RSRP-based cell selection criterion.

\end{document}